\documentclass[journal=nalefd,manuscript=article]{achemso}
\usepackage[version=3]{mhchem} % Formula subscripts using \ce{}

\author{Titus Sandu}
\affiliation{National Institute for Research and Development in Microtechnologies, Bucharest, Romania}
\altaffiliation{International Center for Biodynamics, Bucharest, Romania}
\email{titus.sandu@imt.ro}
\author{Daniel Vrinceanu}
\affiliation{Department of Physics, Texas Southern University, Houston, Texas 77004, USA}
\author{Eugen Gheorghiu}
\affiliation{International Center for Biodynamics, Bucharest, Romania}
\title[An \textsf{achemso} demo]
  {Surface plasmon resonances of clustered nanoparticles}

\abbreviations{IR,NMR,UV}

\begin{document}
%%%%%%%%%%%%%%%%%%%%%%%%%%%%%%%%%%%%%%%%%%%%%%%%%%%%%%%%%%%%%%%%%%%%%
%% The manuscript does not need to include \maketitle, which is
%% executed automatically.  The document should begin with an
%% abstract, if appropriate.  If one is given and should not be, the
%% contents will be gobbled.
%%%%%%%%%%%%%%%%%%%%%%%%%%%%%%%%%%%%%%%%%%%%%%%%%%%%%%%%%%%%%%%%%%%%%
\begin{abstract}
  Linear clusters made by tightly connecting two or more metallic nanoparticles
have new types of surface plasmon resonances as compared to isolated nanoparticles.
These new resonances are related to the size of the junction and to the 
number of interconnected particles and have direct interpretation as
eigenmodes of a Boundary Integral Equation (BIE). This formulation allows
effective separation of geometric and shape contribution from electric properties
of the constituents. Results for particles covered by a thin shell 
are also provided. In addition, the present analysis sheds a new 
light on the interpretation of recent experiments from literature.

\end{abstract}

%section{Introduction}
The interaction of electromagnetic fields with discrete systems ranging from
plasmonic nanoparticles \cite{Ozbay2001,Jain2010} to biological cells \cite{Sandu2010} is governed by their
polarizabilities and is strongly dependent on both dielectric and geometric properties of
particles and surrounding medium. Metallic nanoparticles such as noble-metal
nanostructures exhibit the localized surface plasmon resonance (LSPR)
phenomenon when the electromagnetic radiation interacts with the collective
oscillations of the conduction electrons \cite{Ozbay2001}. The resonances in visible (VIS) and
near-infrarred (NIR) induce extremely strong fields confined to the proximity of the
surface of metallic nanoparticles, and therefore creating the opportunity for highly sensitive
sensors \cite{Lal2007}, surface enhanced Raman scattering (SERS) \cite{Nie1997},
surface enhanced infrared scattering (SEIRS) \cite{Kundu2008},
near-field microscopy \cite{Mock2002}, photoluminescence \cite{Tam2007}, second or higher harmonic
generation \cite{Danckwerts2007,Kim2008}, and other applications. Closely spaced or nearly touching (connected)
nanoparticles \cite{Quinten2001,Atay2004,Romero2006} as well as clustered particles in different configurations
\cite{Le2008} provide additional enhancements of the field within the junction region.
Moreover, the "dumbbell" shaped particles have also shown other remarkable
properties like the large red-shift in the dipolar response \cite{Atay2004,Lassiter2008}, and large
variations in the spectral response \cite{Danckwerts2007,Atay2004}. 

\begin{figure}[b!]
\includegraphics[width=2.5in]{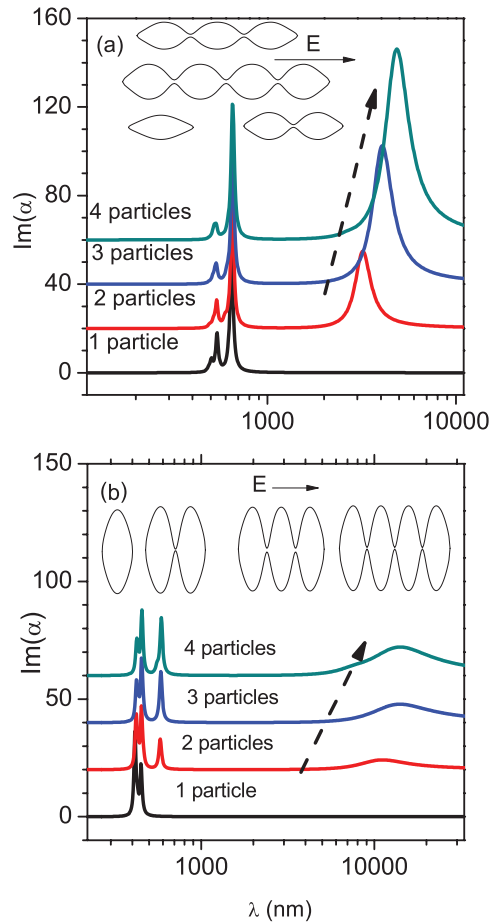}
\caption{\label{fig1}
Imaginary part of polarizabily for clusters made of elongated (a) and
flatten (b) particles in longitudinal fields. Mid-IR resonances are 
associated with the interparticle joints and become more prominent as
the joints tighten. Resonances in the visible spectrum are associated with
the shape and with the number of particles.}
\end{figure}

We demonstrate the spectral BIE approach by calculating the dielectric
response for clusters made of up to four rotationally symmetric particles of
two kinds: elongated particles, longer along the rotation axis (aspect
ratio greater than unity) and flatten particles, shorter along the
rotation axis (aspect ratio less than unity). \ref{fig1} shows the
imaginary part of the polarizability for such clusters, when the electric
field is along the rotation axis. Resonances in the mid-infrared (mid-IR) region of the spectrum
appear for both kinds of clusters and are associated with the joints between the
particles. These resonances are enhanced as the relative
size of the joints become smaller and they  \cite{Sandu2010}. The additional resonance for
clusters of more than one flatten particles (sub-figure b) appears because
the cluster as a whole becomes elongated. This resonance is visible
for clusters made of any number of elongated particles (sub-figure a).
There are no new resonances introduced when the number of particles
increases.
In recent works \cite{Romero2006,Danckwerts2007,Lassiter2008} the distinct optical behavior 
of touching dimers has been attributed to the joints as soon as the particles come 
in conductive contact in the 
dimer \cite{Romero2006}. Our results indicate that the resonances due to the joints are located 
far in mid-IR region of the optical spectrum. On the other hand, in the papers cited above
\cite{Romero2006,Danckwerts2007,Lassiter2008}, the shape of particles used in numerical 
simulations was close to spherical shape, which is the border shape of flatten 
particles. This observation together with the results of \ref{fig1} suggests that 
the red-shifted plasmon resonance observed in \cite{Romero2006,Danckwerts2007,Lassiter2008}
for dimers is the additional resonance which appears in clusters made of flatten particles. 

%section{Results and discussion}
The idea of treating surface plasmons as electrostatic resonances has been
introduced in \cite{Fredkin2003} by showing that source-free electromagnetic fields correspond
to the eigenvalues of the BIE \cite{Mayergoyz2005}. Using a similar approach, we show that a
spectral representation of BIE provide the dielectric responses of homogeneous nanoparticles and
nanoshells in compact forms, which can be directly used in designing plasmonic structures.
The design of analyte sensing is given as an example, where the explicit equations of plasmon
resonances are used to calculate the refractive index sensitivity for each
eigenmode.   

Each eigenmode of the linear response operator has a contribution in the expression of
polarizability proportional with its coupling weight and
inversely proportional with the depolarization factor. Although this property has been demonstrated
for biological cells \cite{Sandu2010}, it is also valid for metallic nano-particles and nanoshells. 
If an eigenvalue has a non-zero weight and is close to the value 1/2,
then it is responsible for a mid-infrared (mid-IR) plasmon resonances,
which can be used in optical nanoantennas \cite{Muhlchlegel2005,Alu2008}. As the interparticle junction tightens,
the eigenvalue approaches 1/2 and the resonance is enhanced.
The coupled-particle LSPR occurs at  a frequency red-shifted from that of a single-particles \cite{Jain2010}.

The nanoparticle clusters are considered under the
electrostatic (quasistatic) approximation \cite{Bohren1998}, when their
size is no more than one tenth of the wavelength of the incoming radiation. 
The optical behavior of metallic nanoparticles \cite{Fredkin2003,Mayergoyz2005} as well as the dielectric
behavior of living cells \cite{Sandu2010}
can therefore be obtained from a solution of the Laplace equation derived from the
corresponding BIE written for the surface $\Sigma$ which defines the particle.
The polarizability of such an object of permittivity $\epsilon_i$
embedded into a medium of permittivity $\epsilon_o$ is given in \cite{Sandu2010,Mayergoyz2005}, and 
has the following spectral representation
\begin{equation}\label{alpha}
\alpha = \sum_k \frac{p_k}{1/(2\lambda) - \chi_k}
\end{equation}
Here electric parameter $\lambda = (\epsilon_i - \epsilon_o)/(\epsilon_i + \epsilon_o)$,
while $\chi_k$ and $p_k$ are
eigenvalues and coupling weights corresponding to the $k$-th
eigenmode of the linear response operator $M$. This geometric operator depends
only on the shape of the particle, is defined by
\begin{equation}\label{M}
M[\mu] = \frac 1{4\pi} \oint_\Sigma 
\frac{{\bf n}({\bf x}) ({\bf x} - {\bf y})}{|{\bf x} - {\bf y}|^3}\; d\Sigma({\bf y})
\end{equation}
and have the following properties \cite{Sandu2010,Fredkin2003}: its spectrum is discrete, real
and bounded to the interval [-1/2, 1/2]; the value 1/2 is always an eigenvalue,
and it has only one active eigenvalue for spherical particles.
The imaginary part of the polarizability $\alpha$ is directly related to optical 
absorption of metallic
nanoparticles \cite{Bohren1998}, and depends 
on the polarization of the incident radiation \cite{Sandu2010,Mayergoyz2005} through dipole-coupling weights $p_k$.
Equation 1 shows how material properties, all collected within $\lambda$, are separated
from geometric properties contained in $\chi_k$ and $p_k$.

An object covered by a very thin shell of permittivity $\epsilon_S$ has a polarizability
calculated from \ref{alpha} by replacing 
$\lambda$ by $\lambda_k = (\tilde\epsilon_k - \epsilon_o)/(\tilde\epsilon_k + \epsilon_o)$
\cite{Sandu2010}, where the effective mode-dependent permittivity
$\tilde\epsilon_k$ is given by:
\begin{equation}\label{tildeEPS}
\tilde \epsilon_k = \epsilon_S \left(
1 + \frac{\epsilon_i - \epsilon_S}{\epsilon_S + \delta (1/2 - \chi_k) 
\epsilon_i + \delta(1/2 + \chi_k) \epsilon_S} \right)
\end{equation}
where $\delta$ is a small parameter roughly equal to the relative volume of the shell with
respect to the particle \cite{Sandu2010}. This
equation allows a direct connection to the hybridization model \cite{Prodan2003}

Within various ranges of frequencies, the dielectric functions 
have in general the form $\epsilon = A+B/W$, where $A$ and $B$ are nearly
constant functions of angular frequency $\omega$, while $W$ is a fast varying function of
$\omega$. In the radiofrequency domain 
$\epsilon = \varepsilon - i \sigma/(\varepsilon_{vac} \omega)$ 
with $\varepsilon$ the static permittivity, 
$\varepsilon_{vac}$ the vacuum permittivity and
$\sigma$ the DC conductivity. In the optical regime dielectrics can be
successfully described by a nearly constant dielectric function
$\epsilon = \varepsilon_d$ and metals
by a Drude dielectric function 
$\epsilon = \varepsilon_m - \omega_p^2/\omega/(\omega + i\gamma)$,
where $\varepsilon_m$ is the interband
contribution, $\omega_P$ is the electron plasma frequency and
$\gamma$ is the damping constant.

The frequency-dependent polarizability of a homogeneous metal nanoparticle of
arbitrary shape can then be written as a sum
of Drude-Lorentz terms by using expansion \ref{alpha}:
\begin{equation}\label{eq3}
\alpha_{plasmon} (\omega) = 
\sum_k \frac{p_k (\varepsilon_m - \varepsilon_{do})}{\varepsilon_{\mbox{\small eff}}} -
\end{equation}
\[
- \frac{p_k}{1/2-\chi_k} \frac{\varepsilon_{do}}{\varepsilon_{\mbox{\small eff}}}
\frac{\tilde\omega_{pk}^2}{\omega(\omega + i\gamma) - \tilde\omega_{pk}^2}
\]
where $\varepsilon_{do}$ is the permittivity of the outer dielectric medium and
the effective dielectric parameter is 
$\varepsilon_{\mbox{\small eff}} = 
(1/2 + \chi_k) \varepsilon_{do} + (1/2 - \chi_k) \varepsilon_{m}$.
For very small particles/clusters dissipation can be neglected ($\gamma \ll \omega_p$) 
and $\tilde\omega_{pk}$ is the frequency of the localized plasmon resonance, which depends
explicitly on the dielectric and geometric parameters of the particle:
$\tilde\omega_{pk}^2 = (1/2 - \chi_k) \omega_p^2/\varepsilon{\mbox{\small eff}}$.

Each active mode in \ref{eq3} has a fast-varying (Drude-Lorentz term)
factor proportional
to its coupling weight $p_k$ and inversely proportional to $(1/2-\chi_k)$, the 
depolarization factor. In clustered particles with tight junctions, several eigenvalues
get closer to 1/2 and the factor $p_k/(1/2-\chi_k)$ has an important contribution even though
the couplings $p_k$ are vanishingly small. The local field enhancement factor
(the height of the resonance peak) is  
$p_k/(1/2 - \chi_k) (\varepsilon_{do}/\varepsilon_{\mbox{\small eff}}) (\tilde\omega_{pk}/\gamma)$
as in the harmonic oscillator model \cite{Bohren1998}.
As $\chi_k$ approaches 1/2 for tighter and tighter junctions the plasmon resonance
red-shifts toward lower frequencies.

The polarizability of dielectric particles of permittivity $\varepsilon_{di}$ covered by
a thin metallic layer and embedded into a dielectric medium of permittivity
$\varepsilon_{do}$ is obtained in the first order in $\delta$,
from \ref{alpha} and \ref{tildeEPS} as
\begin{equation}\label{eq4}
\alpha_{\mbox{shell}} \approx \sum_k \frac{p_k (\varepsilon_{di} - \varepsilon_{do})} 
{\varepsilon_{\mbox{\small eff}}} -
\end{equation}
\[
- \frac{p_k}{1/2-\chi_k} \frac{\varepsilon_{do}}{\varepsilon_{\mbox{eff}}}
\left(\frac{\tilde\omega'^2_{pk}}{\omega (\omega+ i\gamma) - \tilde\omega'^2_{pk}} + 
\right.
\]
\[
\left.
\frac{\tilde\omega''^2_{pk}}{\omega (\omega+ i\gamma) - 
\tilde\omega''^2_{pk} - \omega_p^2/\varepsilon_m}
\right)
\]
with
$\tilde\omega'^2_{pk} = (\delta (1/2+\chi_k)(1/2 - \chi_k) \omega_p^2)/
\varepsilon_{\mbox{eff}}$,
$\tilde\omega''^2_{pk} = (\varepsilon_{di}/\varepsilon_{m})
(\delta(1/2-\chi_k)^2 \omega_p^2)/\varepsilon_{\mbox{eff}}$
and $\varepsilon_{\mbox{eff}} = (1/2 + \chi_k)\varepsilon_{do} + 
(1/2 - \chi_k)\varepsilon_{di}$.
Like the hybridization model \cite{Prodan2003}, \ref{eq4} exhibits two plasmon resonances. 
One resonance frequency for a shelled particle is red-shifted with respect to the
one of the homogenous particle by a factor of $\sqrt{\delta(1/2+\chi_k)}$.
The value of the peak of the resonance is also reduced by the factor 
$\sqrt{\delta(1/2+\chi_k)}$.
Results obtained in \cite{Zhu2008} can be derived from \ref{eq4}, by observing that 
$\varepsilon_{\mbox{\small eff}}$ determines the
influence of $\varepsilon_{di}$ and $\varepsilon_{do}$ on the plasmon resonance in nanoshells.
The other resonance, located at 
$\sqrt{\tilde\omega''^2_{pk} + \omega_p^2/\varepsilon_m}$
is blue-shifted with respect to $\tilde\omega'_{pk}$ and
$\omega_{pk}$ and is smaller than both the red-shifted resonance and the
resonance of a similar homogeneous particle.

A straightforward application of \ref{eq3} and \ref{eq4} is in the
design of analyte sensing.
Large changes in the position of the plasmon resonance in response to small changes in
the refractive index of the outer medium are desired.
The refractive index $n_{do}$ sensitivity for the wavelength $\lambda_{pk}$ of each dipole
active eigenmode is  
$d\lambda_{pk}/dn_{do} = (1/2 + \chi_k) n_{do} \lambda_{pk}/\varepsilon_{\mbox{eff}}$
which increases with increasing wavelenght and
$\chi_k$. A body of theoretical and
experimental research \cite{Jain2010} has suggested that either elongated particles or thin nanoshells
have high refractive index sensitivity. A direct explanation of this derives directly
from \ref{eq3} and \ref{eq4} which show the red shift of the plasmon resonance for tightly joined
clusters, elongated particles and nanoshells \cite{Jain2010,Anker2008}.
A figure of merit is given by the ratio of refractive
index sensitivity over the width of the plasmon resonance.
In the limit $\gamma \ll \omega_p$, the
resonance width is given by the damping constant $\gamma$.
In the case of nanoparticles $\gamma$ is given by the the damping constant of the bulk and an
additional term due to the surface scattering of conduction electrons:
$\gamma = v_F (1/l + 1/L)$ \cite{Bohren1998},
with $v_F$ the Fermi velocity (about $1.4 \times 10^6$ m/s for both Ag and Au),
$l$ the bulk room temperature mean free path (about 52 nm for Ag and 40 nm for Au),
and $L$ is the effective mean free path for collision with the particle's boundary.
For a spherical particle of diameter $d$, $L = 2d/3$, such that if $d$ = 25 $\sim$ 30 nm, 
$\gamma$ is about 0.1 eV.
The damping constants in nanoshells are even greater due to larger
effective mean free paths for collision \cite{Moroz2008}; therefore nanorods are
better than nanoshells when used in analyte sensing.
Using \ref{eq3} and \ref{eq4} the refractive index sensitivity of nanoparticles can be directly
calculated as a function of the aspect ratio or as a function of composition of
the nanoparticle clusters.

We have analyzed clusters of nanoparticles as $n$ touching objects with a rotational
surface generated by the equation:
$x = g(z) \cos\phi$, $y = g(z) \sin\phi$ and
$ -L_n \le z \le L_n$ (the insets of Figure 1a and 1b) with
\begin{eqnarray*}\label{eq6}
g(n,h,A,a,b,z) = 
A\left( 1 + \mbox{sign}\left(z\;\mbox{sign}(z) - (n-1) a
\right) \right) \times \\
\frac{\sqrt{1 - \left(\left(z - \mbox{sign}(z)(n-1)a\right)/a\right)^2}}
{1 + \left(1 + b(z - \mbox{sign}(z)(n-1)a\right)^2} 
\left[ 1 - \mbox{floor}\left( \frac{\mbox{sign}(z)\; z + a}{na}\right)\right]\times\\
\left[ h + 2(A-h) \left( 1 - \frac 1
{1 + (1 - H(z))^2}
\right)\right]
\end{eqnarray*}
where $H(z) = \mbox{mod}((-1)^{\mbox{floor}(z/a)+n-1}\;z, a)^2$,
sign($z$) is signum function, -1, 0 or 1 if z is negative, zero or
positive, floor($z$) is the greatest integer less than $z$, and
mod($x,y$) is the remainder of the division of $x$ by $y$.
Parameters $A$, $a$, $b$ determine the shape of an individual particle,
for example, the ratio $A/a$ is the aspect ratio, and
$h$ determines the size of the connecting gap, and $L_n = na$.

\begin{figure}
\includegraphics[width=2.4in]{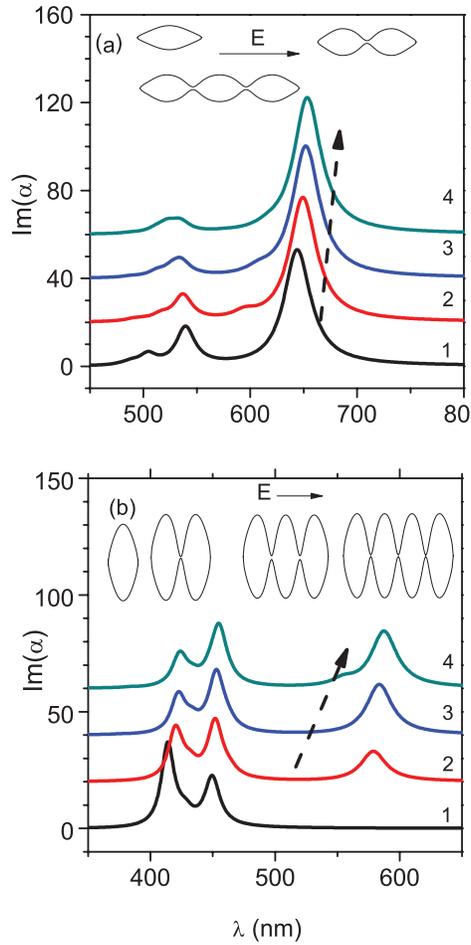}
\caption{\label{fig2}
Imaginary part of polarizabily for clusters of elongated (a)
($A = 2$, $a=3$, $h=0.2$, $b=0.1$) and flatten (b)
($A = 3$, $a=2$, $h=0.05$, $b=0.1$) particles in visible
spectrum.
}
\end{figure}

\ref{fig1} shows the resonances in the mid-IR introduced
by tight junctions because some eigenmodes have eigenvalues
near 1/2 and non-zero, but very small, weights. 
In a recent paper we have shown
that in a cluster of $n$ connected particles there are $n-1$ eigenvalues
close to 1/2 (the smaller the junction the closer the
eigenvalues) \cite{Sandu2010}.
However, only the eigenmodes with antisymmetric
combination of charge distribution in the lobes
are active and contribute to polarizability.
Thus the eigenmodes close to 1/2 induce mid-IR plasmon resonances in optical spectra
of metallic clusters even though the weight (dipole coupling) is quite small. In
our examples clusters of two and three particles have just one resonance, while
for clusters of four particles there are two overlapping resonances. As a rule, all
resonances move toward infrared with increasing number of particles in the clusters,
as seen in \ref{fig2}.

\begin{figure}
\includegraphics[width=2.4in]{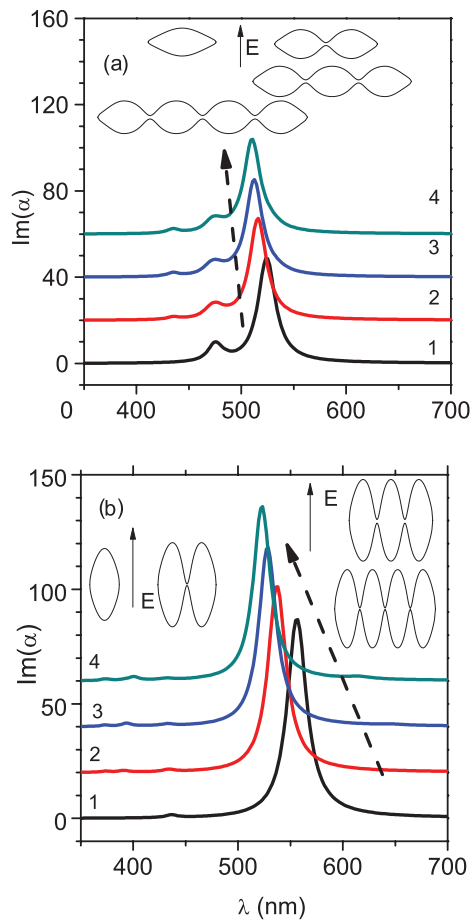}
\caption{\label{fig3}
Imaginary part of polarizabily for clusters of elongated (a)
and flatten (b) particles placed in electric perpendicular to
the rotation axis.
}
\end{figure}

In contrast to parallel fields (\ref{fig1} and \ref{fig2}),
the polarizability for clusters in field perpendicular to their axes,
displays no new significant resonances, as seen in \ref{fig3}.
The existing resonances move toward ultraviolet as $n$ increases.
Also, the mid-IR resonance shifts toward larger wavelengths as the ratio $h/A$ decreases.

\ref{fig2} and \ref{fig3} indicate that, in clusters as well as in individual
nanoparticles, the amplitude of plasmon resonance depends also on the direction
of polarization. Thus flatten particles have the greatest plasmon resonances for
transverse polarization, while elongated particles has theirs for longitudinal fields.
The explanation is based on the harmonic oscillator model, which relates the amplitude of
plasmon resonance to the polarizability of the particles \cite{Bohren1998}.

To conclude this letter, the spectral BIE is a rational approach which provides useful 
insights of the plasmonic behavior of metal nanoparticles.  
Linear clusters of overlapping nanoparticles are good candidates for optical
sub-wavelength guides \cite{Lal2007,Romero2006}.
Our results suggest
that the collective resonances originating from single-particle resonances can
be easily activated by exciting just one particle at one end of the cluster.
In contrast, the collective resonances that are not present in an individual particle
can be successfully activated by exciting the last two particles at the end of the
cluster.
Another possible application of linear clusters of overlapping
nanoparticles is their use as optical nano-antennas instead of nanorods \cite{Muhlchlegel2005,Alu2008}.
The utilization of linear clusters as optical nano-antennas has a lot of promise
due to recent advances in the manufacturing of highly uniform dimer arrays
with gaps smaller than 10 nm \cite{Acimovic2009}. Thus instead of using long and narrow
nanorods, one can use linear clusters, with lengths comparable to the length
of the nanorods and junctions comparable to the cross section of the nanorods \cite{Sandu2010}.

\begin{acknowledgement}
This work has been supported by the Romanian Project ``Ideas'' No.120/2007 and FP 7 Nanomagma No.214107/2008. 
\end{acknowledgement}

%\begin{suppinfo}

%his will usually read something like: ``Experimental procedures and
%haracterization data for all new compounds. The class will
%utomatically add a sentence pointing to the information on-line:

%end{suppinfo}

%\bibliography{achemso-demo}

\end{document}